# Genetic Transferability of Anomalous Irradiation Alterations of Antibiotic Activity

George E. Bass[a,*], Domnita Crisan[b] and Ruxandra E. Alexandrescu[b]

[a] College of Pharmacy, University of Tennessee Health Science Center, Memphis, Tennessee 38163

[b] Postgraduate Department of Biochemistry, Fundeni Hospital Medical School, Bucharest, Romania

[*]Author to whom correspondence should be addressed. Email gbass@utmem.edu

Key Words: irradiation, bacterial growth, chloramphenicol, plasmid, r-factor, acetyl transferase.

**Abstract**

It previously has been discovered that visible light irradiation of crystalline substrates can lead to enhancement of subsequent enzymatic reaction rates as sharply peaked oscillatory functions of irradiation time. The particular activating irradiation times can vary with source of a given enzyme and thus, presumably, its molecular structure. The experiments reported here demonstrate that the potential for this anomalous enzyme reaction rate enhancement can be transferred from one bacterial species to another coincident with transfer of the genetic determinant for the relevant enzyme. In particular, the effect of crystal-irradiated chloramphenicol on growth of bacterial strains in which a transferable R-factor DNA plasmid coding for chloramphenicol resistance was or was not present (*S. panama* $R^+$, *E. coli* $R^+$, and *E. coli* $R^-$) was determined. Chloramphenicol samples irradiated 10, 35 and 60 sec produced increased growth rates (diminished inhibition) for the resistant *S. panama* and *E. coli* strains, while having no such effect on growth rate of the sensitive *E. coli* strain. Consistent with past findings, chloramphenicol samples irradiated 5, 30 and 55 sec produced decreased growth rates (increased inhibition) for all three strains.

**Introduction**

What has evolved as a possibly universal capability on the part of enzymes to recognize unorthodox radiation signals was first reported in 1968 (Comorosan et al., 1968, 1970a,b,c, 1971a,b, 1973, 1975, 1980). The study presented here examines the genetic origin of the relevant enzyme characteristic by utilizing inter-bacterial transfer of an R-factor plasmid. The particular plasmid chosen codes for production of the enzyme chloramphenicol acetyl transferase (CAT) which, in turn, is responsible for bacterial resistance to chloramphenicol (Smith, et al., 1972).

Comorosan and co-workers have studied extensively the subject phenomenon. Typically, alteration of *in vitro* activity of an enzyme (Comorosan, et al., 1972b) or growth rate of a microorganism (Comorosan et al., 1973) is obtained as a consequence of visible light irradiation of the enzyme substrate or microorganism growth factor (or inhibitor) in the crystalline state. That is, irradiation is performed prior to dissolution and introduction of the substrate or growth factor into the reaction mixture or growth medium. Probably the most unexpected feature of these observations is that the activity alteration is a periodic function of the duration of exposure of the crystalline substrate to the electromagnetic radiation employed. Thus, if $t_m$ is the shortest irradiation time which produces an alteration, the next alteration will occur with irradiation for an increased length of time, $t_m + \tau$. The entire collection of irradiation times which produce activity alteration, $t^*$, can be represented as $t^* = t_m + n\tau$ where n is an integer and $\tau$ a constant. The magnitudes of $t_m$ and $\tau$ are on the order of seconds (5-45 seconds) while the width of the activation peaks are $\leq 0.5$ seconds. Irradiations for other times $t \neq t^*$, whether longer or shorter, produce no activity alteration relative to non-irradiated controls. All $t_m$ and $\tau$ values have been found to be multiples of 5 seconds. These basic observations have been reproduced or demonstrated successfully in other laboratories (Bass, et al., 1973, 1976a,b, 1977; Etzler and Westbrook, 1986; Goodwin and Vieru, 1975; Sherman, et al., 1973, 1974).

The mechanism by which the altered biological responses are produced remains unestablished. For convenience, the responses and their associated irradiation times, $t^*$, are referred to hereafter as "signals."

In vitro studies (Comorosan, et al., 1971a,b) with systems of enzymes from the glycolysis, gluconeogenesis, and Krebs cycle pathways have revealed patterns in the $t^*$ values for the individual reactions. These patterns have led to the suggestion that this phenomenon reflects an innate ability of enzymes to recognize, or respond to, signals which are partly responsible for, or associated with, cellular metabolic control. Further, the $t^*$ values for a particular enzyme isolated from mammalian tissues are found to be numerically different from those for the corresponding enzyme from a microorganism. Thus, there is a species dependence which implies a dependence on composition or substructure of the enzyme molecule. Additional support for this proposition has been obtained with isoenzyme rehybridization studies (Comorosan, et al., 1972a).

This strange behavior observed in the kinetics of isolated, purified enzymes is also found, at least qualitatively, in studies on growth rate of microorganisms (Comorosan, et al., 1973, 1975; Bass and Crisan, 1975; Sherman, et al., 1974). Thus, t* irradiation times corresponding to growth rate enhancements have been found for introduction of crystal-irradiated arginine, histidine, and tryptophan into minimal media for growth of yeast strains auxotrophic for these amino acids. Presumably, these growth rate alternations are a consequence of stimulation of some particular enzyme reaction which utilizes the irradiated amino acid. The studies to be described here speak to this presumption.

Corresponding to microbial growth rate enhancements with irradiated growth factors, increased inhibition of growth is obtained for the action of crystal-irradiated antibiotics on sensitive yeast and bacteria (Comorosan, et al., 1975; Bass and Crisan, 1975; Sherman, et al., 1974). Again, the familiar $t^* = t_m + n\tau$ irradiation time dependence is encountered. Presumably, the irradiation process creates photoproducts which stimulate interaction of the antibiotic with cellular receptors (possibly the ultimate target of the antibiotic such as the ribosomes for tetracycline or chloramphenicol). On the other hand, for antibiotic resistant bacteria, two sets of growth alteration antibiotic-irradiation signals, $t^*_1$ and $t^*_2$, may be found. (As used here, "resistant" bacteria are those which have the capability to actively and appreciably deactivate given antibiotics enzymatically). One set of signals corresponds to increased inhibition of growth (analogous to findings with sensitive microorganisms) while the other set of signals are associated with decreased inhibition of growth (i.e., increased growth). Presumably, these latter signals correspond to antibiotic crystal-irradiation induced enhancement of interaction of the antibiotic with the deactivating enzyme present in the resistant species. Investigation of this point was the primary objective of the study described here.

Good evidence that the "resistance signals" can be attributed to the antibiotic deactivating enzyme has been reported previously for penicillin resistant *B. subtilis* (Sherman, et al., 1974). Identical t* times were found for decreased growth inhibition by crystal-irradiated penicillin and, correspondingly, increased *in vitro* activity of purified penicillinase isolated from two different sources. This brings us finally to the study at hand.

If the resistance signals are associated with stimulation of the antibiotic deactivating enzyme and the particular numerical values dictated by the amino acid sequence of that enzyme, this "information" must also reside in the DNA code for that enzyme. As a first assumption, the *in vivo* structure of that enzyme should be otherwise independent of the particular cellular environment in which it finds itself. Thus, if the corresponding DNA code is transferred to another organism which previously lacked it, a new set of resistance signals should appear for that organism where there were none before. This is the essence of the experiments reported here.

**Materials and Methods**

The antibiotic selected for study was chloramphenicol hemisuccinate (CM). Bacteria employed were (1) a CM sensitive *E. coli* K12 Hfr Cavalli, met$^-$, a competent R-factor receptor (designated hereafter as *E. coli* R$^-$), (2) a *Salmonella panama* (1467) carrying R-factor resistance to CM, tetracycline, ampicillin, streptomycin, sulfathiazole and neomycin (designated hereafter as *S. panama* R$^+$), and (3) the above *E. coli* to which the R-factor carried by *S. panama* R$^+$ had been transferred rendering it resistant to the above antibiotics (designated hereafter as *E. coli* R$^+$). Stock cultures of the three organisms were maintained on nutrient agar (Difco) slants. Antibiotic growth inhibition was assayed turbidimetrically in liquid medium (penassay broth, Difco; Bauch and Lomb Spectronic 20, 540 nm). The inoculum for each morning was prepared the previous afternoon by loop transfer from the appropriate slant to 25 ml penassay broth, incubated overnight at 37 C and diluted with sterile broth such that 0.25 ml placed in 3.00 ml sterile medium produced an optical density reading in the range 0.03 to 0.06. Dose-response curves for CM (Intreprindevea de Antibiotice Iasi, Romania) with each of the three strains were obtained. For each strain, a CM concentration falling on the steep portion of the curve near the ID$_{50}$ point was chosen for the irradiation experiments. These are, for the *S. panama* R$^+$, 1.25 mg/ml broth; for the *E. coli* R$^-$, 8.5 mcg/ml broth; and for the *E. coli* R$^+$, 0.100 mg/ml broth. In the irradiation experiments, these concentrations were obtained by weighing CM samples of 24.4 mg, 12.8 mg and 24.4 mg, respectively. For the *S. panama* R$^+$ studies, the 24.4 mg samples were subsequently dissolved in 1.50 ml sterile water and 0.25 ml of this solution added to 3.00 ml broth. For the *E. coli* R$^-$, the 12.8 mg samples were dissolved in 10.0 ml sterile water and 0.020 ml added to 3.00 ml broth. For the *E. coli* R$^+$, the 13.0 mg samples were dissolved in 10 ml sterile water and .0.25 ml added to 3.00 ml broth.

For each organism, samples containing the above indicated quantities of dry, finely powdered CM were weighed and evenly spread in flat bottom plastic cups of 25 ml capacity. These samples were placed in a desiccator and used within 18 hr. of preparation. In the irradiation experiments, CM samples were assayed in groups of three; each such group is hereafter referred to as a "Run." One of the three CM samples in each Run was not irradiated and served as the control for that Run. For each Run, two of the CM samples were irradiated for specific times, then all three samples dissolved in sterile water and the appropriate quantity (preceding paragraph) added to 3.00 ml penassay broth in 11 x 100 mm colorimeter cuvettes, five cuvettes per CM sample. To each cuvette was than added 0.25 ml of the diluted inoculum and the initial optical densities measured. Optical density readings were then taken hourly until the beginning of the stationary phase was reached. For the *S. panama* R$^+$ and *E. coli* R$^-$ this was 4 hrs.; for the *E. coli* R$^+$, 6 hrs. Thus for a given CM sample irradiated t seconds, the average growth for 5 replicates is obtained as the average increase in optical density over the growth period (either 4 or 6 hrs. as indicated above) and will be designated G(t). The net growth attributed to CM irradiation is then G(t) - G(0), where G(0) is the average growth obtained for the non-irradiated CM sample in the same Run.

Radiation source employed was a high pressure Hg lamp (70w, 220V, Astralux, Wein, bulb ASH250). A narrow band in the vicinity of 546 nm was selected with an optical filter ( λmax = 546.1 nm, band width = 8.8 nm, Ditric Optics 2-cavity bandpass filter). Irradiation time was controlled by an electronic shutter device (Uniblitz, model 100-2, VA). Luminosity at the sample was in the range 440-460 footcandles (Panlux Electronic Footcandle Meter). The lamp was placed with the bulb 17 cm directly above the sample.

**Results**

The first criterion for this study was that irradiated CM could elicit two sets of signals, one of which would correspond to diminished growth inhibition (i.e., resistance signals) in the $R^+$ salmonella strain. That this first requirement was indeed satisfied is demonstrated by the results displayed in Figure 1 (where each point represents net growth attributable to irradiation of CM, see earlier definition of net growth). To provide an indication of the precision of the measurements, vertical bars are placed symmetrically about each point to indicate ± one standard deviation for the associated G(t). As can be seen, decreased growth (i.e., increased inhibition) is obtained for irradiation times of 5, 30 and 55 seconds. These will be referred to as the "sensitive" signals, $t_m$ = 5 sec, $\tau$ =25 sec, which we presume reflect irradiation enhanced interaction of the antibiotic with the ribosomes. On the other hand, increased growth is found for irradiation times of 10, 35 and 60 seconds. These are the "resistance" signals; $t_m$=10 sec and $\tau$ =25 sec. They are presumably due to irradiation induced enhancement of CM deactivation by the enzyme chloramphenicol acetyl transferase which is coded for by the R-factor. The question at hand is whether or not these signals can be transferred to another bacteria (*E. coli* $R^-$) along with the R-factor.

A second criterion is then that the *E. coli* R$^-$ not display a set of signals corresponding to increased growth (i.e., "resistance" signals). The consequences of irradiation of crystalline CM on growth of this organism are displayed in Figure 2. As expected, no resistance signals were observed. The sensitive signals occurred at the same t* times found with the salmonella. While our attention is focused primarily on the resistance signals, presence here of the sensitive signals serves the important role of establishing that the microorganism and procedure are suitable for observation of the phenomenon. With the stage thus set, response of the *E. coli* R$^+$ to crystal-irradiated CM was determined. The results are presented in Figure 3. Found are decreased growth for t* = $t_m$ =5 sec, $\tau$ =25 sec and, where there were no signals before, increased growth for $t_m$=10 sec, $\tau$ =25 sec. Thus, coincident with transfer of the genetic machinery for CM resistance from the *S. panama* R$^+$ to the *E. coli* is the appearance in the latter of resistance signals at exactly the same crystal-irradiation times displayed by the *S. panama* R$^+$.

**Discussion**

Prior experience with other systems led to the hypotheses that (1) these crystal-irradiation signals correspond to specific *in vitro* enzyme reaction rate enhancements, (2) the parallel effect found for microorganism growth rates is due to such specific enzyme reaction rate enhancements, (3) the particular activating irradiation times are determined by the given substrate and the given enzyme (sensibly independent of environment), and (4) that transfer of the genetic determinant for the given enzyme to a different environment (here, a different bacterial species) will not alter the signals (irradiation times) which stimulate its interaction with the given substrate. The experiments reported here were designed to test these working assumptions and the results are found to be entirely consistent therewith. A possible explanation of the phenomenon in terms of photo-driven oscillatory reactions involving the atmospheric gases at the crystalline surface has been proposed (Bass, 2005, 2007).

Acknowledgement: G.E.B. would like to express his appreciation to the Romanian government and Professor Sorin Comorosan for a Visiting Scientist grant and to the University of Tennessee for off-campus assignment during which this study was conducted at Fundeni Hospital Medical School in Bucharest.

Smith, D.H., Harwood, J.H. and Rubin, F.A. (1972). Studies on the regulation of the chloramphenicol acetyl transferase mediated by R factors. In: *Bacterial Plasmids and Antibiotic Resistance*, (V. Kremery, L. Rosival and T. Watanabe, eds.) p.319 Berlin, Springer Verlag.

Fig. 1. Net growth alteration of *S. panama* R[+] attributable to prior irradiation of crystalline chloramphenicol (CM). Net growth = G(t) - G(0), where G(t) is the averaged increase in optical density at 540 nm over a 4 hr. period for 5 replicate cultures containing CM irradiated t sec, and G(0) is the averaged change in optical density under identical conditions with non-irradiated CM. Vertical bars indicate ± one std. dev. for G(t) for the net growth value about which they are symmetrically placed. Average over all G equals 0.391.

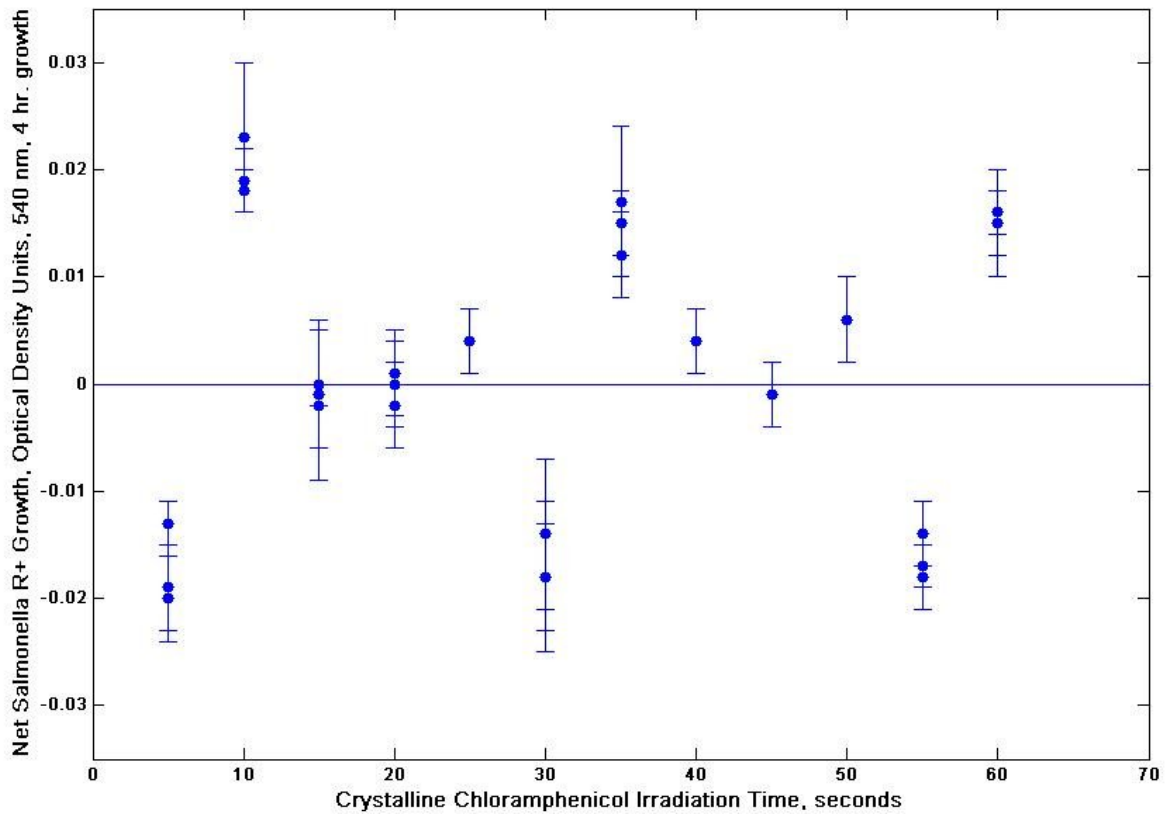

Fig. 2. Net growth alteration of *E. coli* R⁻ attributable to prior irradiation of crystalline chloramphenicol. Net growth defined in Fig, 1. Average over all G equals 0.247.

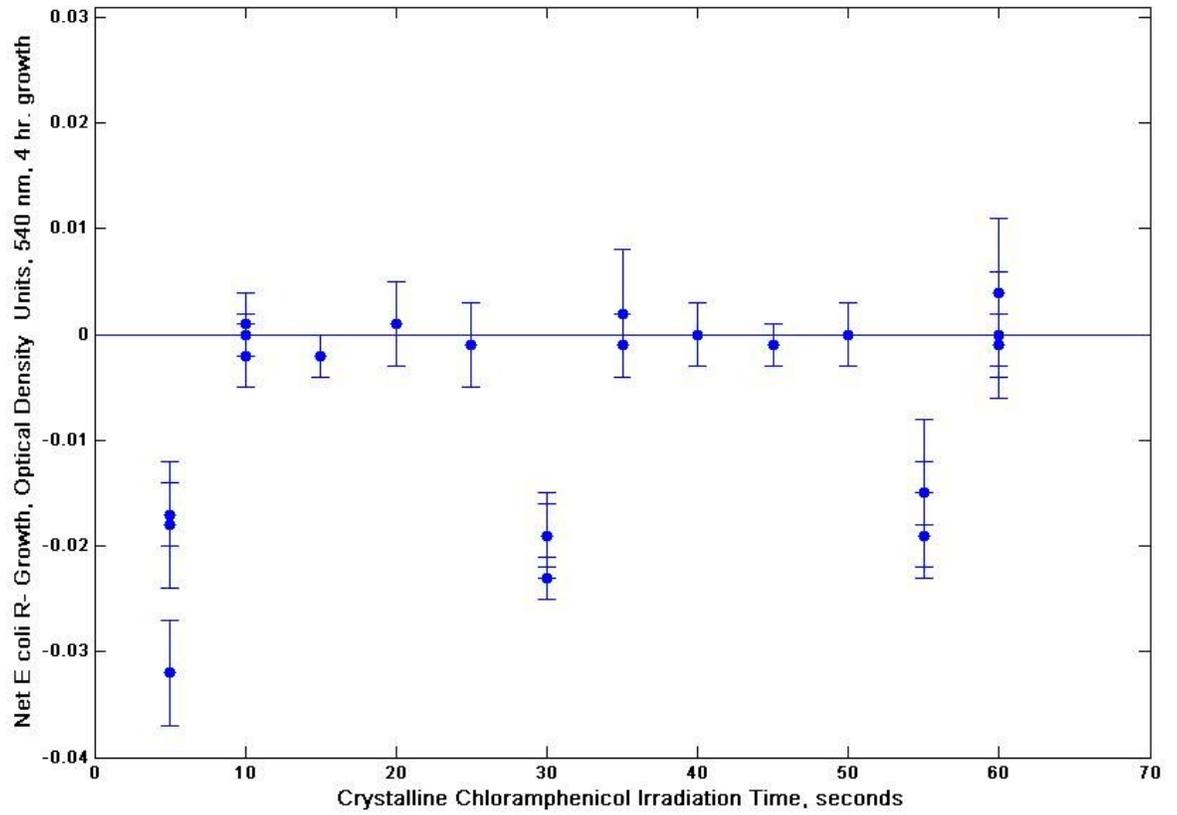

Fig. 3. Net growth alteration of *E. coli* R$^+$ attributable to prior irradiation of crystalline chloramphenicol. Net growth defined as in Fig. 1. except growth period was 6 hr. Average over all G equals 0.231.

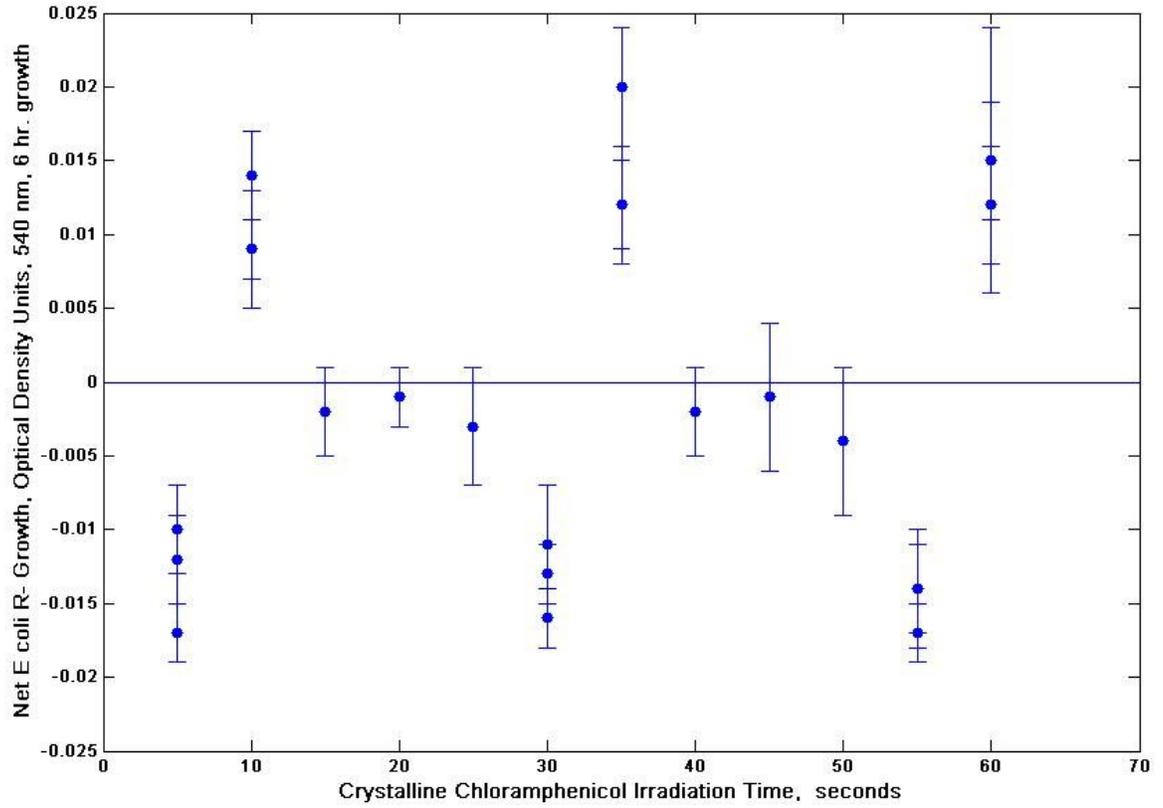